\documentclass[10pt]{article}
\usepackage{amsmath, natbib}
\usepackage{graphicx}
\usepackage{authblk}
\usepackage[top=4cm, bottom=4cm, left=5cm, right=5cm]{geometry}
\usepackage{fancyhdr}
\renewenvironment{abstract}{%
\hfill\begin{minipage}{0.95\textwidth}
\rule{\textwidth}{1pt}}
{\par\noindent\rule{\textwidth}{1pt}\end{minipage}}
\makeatletter
\renewcommand\@maketitle{%
\hfill
\begin{minipage}{0.95\textwidth}
\vskip 2em
\let\footnote\thanks 
{\LARGE\bf \@title \par }
\vskip 1.5em
{\large \@author \par}
\end{minipage}
\vskip 1em \par
}
\makeatother
\begin{document}
%
\title{OVI 6830\AA\ Imaging Polarimetry of Symbiotic Stars}
\author[1]{Akras Stavros}
\affil[1]{Observat\'{o}rio Nacional/MCTIC, Rio de Janeiro, Brazil}
%
\maketitle

\begin{abstract}
I present here the first results from an ongoing pilot project with the 1.6~m telescope at the OPD, Brasil, aimed at the 
detection of the OVI $\lambda$6830 line via linear polarization in symbiotic stars. The main goal is to demonstrate that OVI 
imaging polarimetry is an efficient technique for discovering new symbiotic stars. The OVI $\lambda$6830 line is found in 
5 out of 9 known symbiotic stars, in which the OVI line has already been spectroscopically confirmed, with at least 3$\sigma$ 
detection.  Three new symbiotic star candidates have also been found.
\end{abstract}


\section{Introdution}

Raman scattering is a well established mechanism in symbiotic stars (Schmid 1989; Nussbaumer et al. 1989). Generally, 
two broad lines are detected in symbiotic stars (SySts) centred at 6830\AA\ and 7088\AA, with the latter being approximately 
4 times weaker. These two lines are attributed to Raman scattering of the ultraviolet OVI $\lambda\lambda$1032, 1038 
resonance lines by neutral hydrogen (Schmid 1989; Nussbaumer et al. 1989). Of the confirmed Galactic SySts 252 or 55\% 
(Akras et al. these proceedings) show the broad OVI $\lambda$6830 Raman-scattered line (Allen 1980; Schmid \& Schild 1994). 

The mechanism of Raman scattering is well known to produce strong polarization from a few percent up to 10-15\% 
(Schmid \& Schild 1994; Harries \& Howarth 1996,2000). Spectro-polarimetric observations are very important for 
determining the orbital parameters for these systems as well as the mass-loss rate of the cold giant by studying the O~VI 
line profiles (Harries \& Howarth 1997). However, only a few of them have been systematically observed.  Hence, an 
OVI $\lambda$6830 imaging polarimetric survey of SySts is required in order to unveil those with significant polarization. 

In this paper, I present the first OVI 6830\AA\ imaging polarimetric observations of SySts. This pilot project aims to pave the way 
for larger and systematic OVI surveys (imaging photometry and/or polarimetry) searching for new SySts without necessarily obtain 
follow-up observations. 

\section{Observations}
Linear imaging polarimetric observations were obtained with the 1.6~$\mu$m Perkin-Elmer telescope at the Observat\'{o}rio do Pico dos Dias 
(OPD/LNA) in Brasil. The data were taken using the imaging polarimeter Instituto de Astronom\'{i}a, Geof\'{i}sica
e Ci\^{e}ncias Atmosf\'{e}ricas polarimeter (IAGPOL; Magalh\~{a}es et al. 1996). The polarimeter consists of a half-wave plate
that can be rotated in steps of 22.5 degrees and a Savart calcite prism (Ram\'{i}rez et al. 2015). The broad-band filter {\it R} and a 
narrowband filter centred at 6810\AA\ (100\AA\ bandwidth) were used with exposure times between 5 and 240~s. The field of view and
image scale are 12 arcmin$^2$ and 0.36 arcsec pixel$^{-1}$, respectively. During observations, the seeing varied between 1.5 and 2 arcsec.
Fig.~1 shows an image of RR Tel obtained with the OVI $\lambda$ filter.   

Separate images were obtained for both filters and at relative position angles of the prism of 0, 22.5, 45, 67.5, 90,
112.5, 135, 157.5 and 180 degrees. The instrumental polarization as well as the rotation angle of the polarimeter were estimated
by observing a number of unpolarized and polarized standard stars during the campaign. The data reduction was performed 
by using the IRAF package BEACON pipeline. The code applies a standard reduction technique which includes, removal 
of cosmic rays, subtraction of the dark current and bias, as well as flat-field normalization. It calculates the degree of 
polarization (DoP) and position angle (PA) for several stars in the field.

\begin{figure}
\centering
\includegraphics[width=8cm]{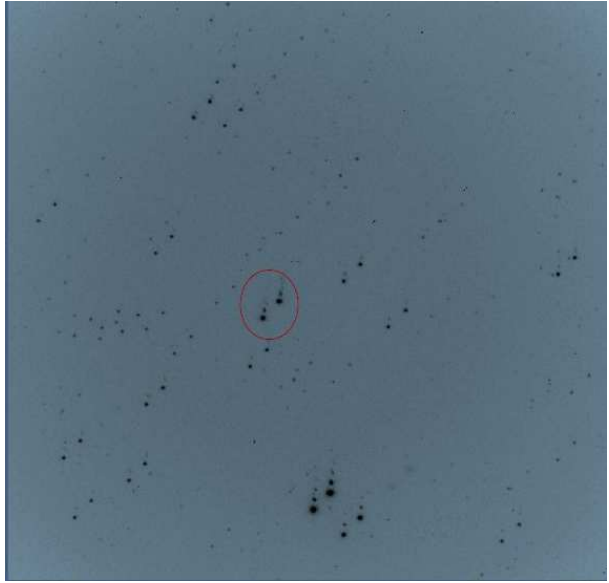}
\caption {OVI $\lambda$6830 line image of RR Tel taken with the IAGPOL at position angle of the prism of 0 degree. 
The two stars in the red circle refer to the real and imaginary part of RR Tel. North is up, east to the left.}
\label{}
\end{figure}

\begin{table*}
\caption[]{Degree of linear polarization of SySts.}
\label{table5}
\begin{tabular}{llllllllllll}
\hline 
Name       & O~VI$^{\dag}$  & P$_{\rm OVI}$ & P$_{\rm R}$  & P$_{\rm ISM,OVI}$ & P$_{\rm ISM,R}$\\
           & (6830\AA)      & (\%)           &  (\%)       & (\%)               & (\%)  \\ 
\hline
{\it RR Tel}     &  yes        & 3.09$\pm$0.13 & 0.58$\pm$0.02& 0.42$\pm$0.03 &  0.45$\pm$0.02\\
{\it Hen 2-106}  &  yes        & 4.19$\pm$0.20 & 3.02$\pm$0.03& 2.21$\pm$0.02 &  2.13$\pm$0.01\\
{\it CD-43 14304}&  yes        & 0.65$\pm$0.03 & 0.22$\pm$0.01& 0.53$\pm$0.02 &  0.28$\pm$0.01\\
{\it AR PAV}     &  yes        & 1.30$\pm$0.03 & 1.16$\pm$0.02& 0.61$\pm$0.01 &  0.61$\pm$0.01\\
{\it 2MASS16422739}&  yes      & 9.32$\pm$3.07 & 4.79$\pm$0.06& 1.91$\pm$0.01 &  1.71$\pm$0.01\\
\hline
BI Cru     &  no         & 1.04$\pm$0.03 & 1.07$\pm$0.06& 1.33$\pm$0.02 &  1.26$\pm$0.02\\
HD 330036  &  no         & 2.79$\pm$0.06 & 2.66$\pm$0.02& 2.06$\pm$0.02 &  2.01$\pm$0.01\\
Hen 3-1213 &  no         & 2.82$\pm$0.20 & 3.04$\pm$0.14& 2.01$\pm$0.01 &  2.10$\pm$0.01\\
Hen 3-1761 &  no         & 1.00$\pm$0.02 & 1.01$\pm$0.02& 0.97$\pm$0.01 &  0.98$\pm$0.01\\
V4018      &  no         & 0.35$\pm$0.03 & 0.40$\pm$0.02& 0.50$\pm$0.01 &  0.51$\pm$0.01\\
V4074      &  yes        & 1.15$\pm$0.07 & 1.05$\pm$0.33& 1.18$\pm$0.02 &  1.25$\pm$0.01\\
Hen 3-1341 &  yes        & 1.89$\pm$0.05 & 1.75$\pm$0.14& 2.01$\pm$0.01 &  2.04$\pm$0.01\\
StHa 164   &  yes        & 2.47$\pm$0.13 & 2.58$\pm$0.17& 2.58$\pm$0.01 &  2.66$\pm$0.02\\
PN~K~3-12  &  yes        & 0.95$\pm$0.17 & 0.75$\pm$0.13& 0.92$\pm$0.02 &  0.93$\pm$0.01\\
\hline
\end{tabular}
\medskip{}
\begin{flushleft}
$^{\dag}$ Spectroscopic detection of the OVI $\lambda$ 6830 Raman-scattered line. 
\end{flushleft}
\end{table*}

\section{Results}
In narrowband polarization imaging there are three components that have to been taken into account, (i) the polarization in spectral line, 
(ii) the polarization in the continuum and (iii) the interstellar medium polarization (ISP). For a 
positive detection of the spectral line via polarization imaging, the continuum and ISP components have to be measured and 
properly subtracted. For the continuum polarization, the broadband filter {\it R} was used. 

Regarding the ISP, the field star method was used (e.g. Akras et al. 2017). Assuming that the field stars are 
unpolarized, the average DoP from all the stars in the field gives a rough estimate of the ISP.

Table~1 presents the DoP in the OVI and {\it R} filters for 
14 known SySts (Belczy\'{n}ski et al. 2000, Akras et al. these proceedings) as well as the ISP
in the direction of each object (5$^{\rm th}$ and 6$^{\rm th}$ columns). Two criteria were used in order to get a positive detection of the 
OVI Raman line, 
(i) P$_{\rm OVI}>$P$_{\rm ISM,OVI}$+3$\times\sigma_{P_{\rm OVI}}$, which implies the star is intrinsically polarized in the OVI line and 
(ii) P$_{\rm OVI}>$P$_{\rm R}$+3$\times\sigma_{P_{\rm OVI}}$, which implies that the observed polarization is due to the spectral line and not the continuum. In some cases, the target can be highly polarized in the continuum ({\it R}) but not necessarily in the spectral line (e.g. Hen 3-1213).

In my sample, there are nine SySts in which the presence of OVI $\lambda$6830 Raman line had been confirmed spectroscopically before 
and five SySts without a confirmed detection. I get a 3$\sigma$ detection for 5 ({\it italic} in Table~1) out of 9 SySts with previous OVI detection. The non detection of the OVI line from the last four SySts may be due to the variation of the line (very faint OVI line at the time of my observations). As for the five SySts, without a confirmed detection, I get no detection for all of them (100\% success).  

Besides the SySts, I have also found three stars that pass the aforementioned criteria and I have considered them as candidate SySts. 
Since this is a pilot project, follow-up spectroscopic observations of these three candidates are required in order to unveil their 
nature and confirm the power of this technique.

\section{Conclusion}
OVI 6830\AA\ imaging polarimetry seems to be very promising technique for searching new SySts.
5 out of 9 SySts (56\%) show a 3 $\sigma$ detection of the OVI Raman line. The detection of the OVI Raman line via 
polarimetry assures the symbiotic nature of the object without following-up observations. 
More observation of SySts, with or without a confirmed detection of the OVI line, are planned to be obtained this year.

\vskip10pt
{\bf\large Acknowledgements}
I would like to thank the organizing committee for the opportunity to attend this event and present this work as 
well as for their financial support. I also acknowledge support of CNPq, Conselho Nacional de Desenvolvimento Cient\'ifico e 
Tecnol\'ogico - Brazil (grant 300336/2016-0). This work is based upon observations carried out at the Observat\'{o}rio do Pico 
dos Dias (LNAMCT/CNPq, Braz\'{o}polis, Brazi). Many thanks also to Nikos Nanouris for reviewing this paper.

\end{document}